\documentclass[twocolumn,showpacs,preprintnumbers,amsmath,amssymb,prb,superscriptaddress]{revtex4}
\usepackage{graphicx}
\usepackage{dcolumn}
\usepackage{subfigure}
\usepackage{bm}
\usepackage{float}
\usepackage{amssymb}
\usepackage{amsmath}
\begin{document}

\title{Pressure-induced enhancement of superconductivity and suppression of semiconducting behavior in the \textit{Ln}O$_{0.5}$F$_{0.5}$BiS$_2$ (\textit{Ln} = La, Ce) compounds}

\author{C. T. Wolowiec}
\affiliation{Department of Physics, University of California, San Diego, La Jolla, California 92093, USA}
\affiliation{Center for Advanced Nanoscience, University of California, San Diego, La Jolla, California 92093, USA}

\author{D. Yazici}
\affiliation{Department of Physics, University of California, San Diego, La Jolla, California 92093, USA}
\affiliation{Center for Advanced Nanoscience, University of California, San Diego, La Jolla, California 92093, USA}

\author{B. D. White}
\affiliation{Department of Physics, University of California, San Diego, La Jolla, California 92093, USA}
\affiliation{Center for Advanced Nanoscience, University of California, San Diego, La Jolla, California 92093, USA}

\author{K. Huang}
\affiliation{Department of Physics, University of California, San Diego, La Jolla, California 92093, USA}
\affiliation{Center for Advanced Nanoscience, University of California, San Diego, La Jolla, California 92093, USA}
\affiliation{Materials Science and Engineering Program, University of California, San Diego, La Jolla, California 92093, USA}

\author{M. B. Maple}
\email[Corresponding Author: ]{mbmaple@ucsd.edu}
\affiliation{Department of Physics, University of California, San Diego, La Jolla, California 92093, USA}
\affiliation{Center for Advanced Nanoscience, University of California, San Diego, La Jolla, California 92093, USA}
\affiliation{Materials Science and Engineering Program, University of California, San Diego, La Jolla, California 92093, USA}

\begin{abstract}

Electrical resistivity measurements as a function of temperature between 1 K and 300 K were performed at various pressures up to 3 GPa on the superconducting layered compounds \textit{Ln}O$_{0.5}$F$_{0.5}$BiS$_2$ (\textit{Ln} = La, Ce).  At atmospheric pressure, LaO$_{0.5}$F$_{0.5}$BiS$_{2}$ and CeO$_{0.5}$F$_{0.5}$BiS$_{2}$ have superconducting critical temperatures, \textit{$T_c$},  of 3.3 K and 2.3 K, respectively.  For both compounds, the superconducting critical temperature \textit{$T_c$} initially increases, reaches a maximum value of 10.1 K for LaO$_{0.5}$F$_{0.5}$BiS$_{2}$ and 6.7 K for CeO$_{0.5}$F$_{0.5}$BiS$_{2}$, and then gradually decreases with increasing pressure.  Both samples also exhibit transient behavior in the region between the lower \textit{$T_c$} phase near atmospheric pressure and the higher \textit{$T_c$} phase. This region is characterized by a broadening of the superconducting transition, in which \textit{$T_c$} and the transition width $\Delta$\textit{$T_c$} are reversible with increasing and decreasing pressure. There is also an appreciable pressure-induced and hysteretic suppression of semiconducting behavior up to the pressure at which the maximum value of \textit{$T_c$} is found. At pressures above the value at which the maximum in \textit{$T_c$} occurs, there is a gradual decrease of \textit{$T_c$} and further suppression of the semiconducting behavior with pressure, both of which are reversible.
\end{abstract}

  % pressure effects on crystal structure 61.50.Ks; transport in in superconductors; 74.25.F-; pressure effects on superconducting transition temperature 74.62.Fj;  non-cuprate materials multinary compounds, 74.70.Dd;

\maketitle

\section{INTRODUCTION}

Superconductivity with a superconducting critical temperature \textit{$T_c$} = 8.6 K has recently been reported in the layered compound Bi$_4$O$_4$S$_3$. \cite{Mizuguchi1, Singh} Following this report, other BiS$_2$-based superconductors including $Ln$O$_{1-x}$F$_x$BiS$_2$ (\emph{Ln} = La, Ce, Pr, Nd, Yb) with a \textit{$T_c$} as high as 10 K have been synthesized and studied.\cite{Li, Jha, Jha1, Deguchi, Kotegawa, Awana, Demura, Mizuguchi2, Xing, Yazici}  More recent work demonstrates that chemical substitution of the tetravalent ions Th$^{+4}$, Hf$^{+4}$, Zr$^{+4}$ and Ti$^{+4}$ for trivalent lanthanum, La$^{+3}$, in LaOBiS$_2$ increases the charge-carrier density and induces superconductivity.\cite{Yazici2} Most of the research on the layered BiS$_2$ compounds has heretofore centered on the effect of chemical substitution on superconductivity. Application of an external pressure may also be employed as a method for reducing the unit cell volume of these compounds and studying the resultant effect on superconductivity. In this paper, we report measurements of the pressure dependence of the normal state electrical resistivity between 1 K and 300 K and \textit{$T_c$} at various pressures up to $\sim$ 3 GPa for the compounds LaO$_{0.5}$F$_{0.5}$BiS$_{2}$ and CeO$_{0.5}$F$_{0.5}$BiS$_{2}$. We compare our results to recently reported studies of LaO$_{0.5}$F$_{0.5}$BiS$_{2}$ samples synthesized under high pressure by Kotegawa \textit{et al.} \cite{Kotegawa}\\

The qualitative evolution of \textit{$T_c$} with pressure is markedly similar for both LaO$_{0.5}$F$_{0.5}$BiS$_{2}$ and CeO$_{0.5}$F$_{0.5}$BiS$_{2}$, which have \textit{$T_c$} values (at atmospheric pressure) of 3.3 K and 2.2 K, respectively. For both compounds, \textit{$T_c$} initially increases, reaches a maximum value of 10.1 K at $\sim$ 1 GPa for LaO$_{0.5}$F$_{0.5}$BiS$_{2}$  and 6.7 K at $\sim$ 2 GPa for CeO$_{0.5}$F$_{0.5}$BiS$_{2}$, and then gradually decreases with increasing pressure. Both compounds also exhibit striking transient behavior in the region between the lower \textit{$T_c$} phase near atmospheric pressure and the higher \textit{$T_c$} phase. This transient region is characterized by a rapid increase of \textit{$T_c$} and an increase of the superconducting transition width $\Delta$\textit{$T_c$}, in which both \textit{$T_c$} and  $\Delta$\textit{$T_c$}  are reversible with increasing and decreasing pressure cycles. This occurs over a range in pressure from $\sim$ 0.5 GPa to 1.1 GPa for LaO$_{0.5}$F$_{0.5}$BiS$_{2}$ and from  $\sim$ 0.5 GPa to 1.5 GPa for CeO$_{0.5}$F$_{0.5}$BiS$_{2}$.  In both materials, there is a sizable pressure-induced suppression of semiconducting behavior exhibiting hysteresis up to the pressure at which the maximum value of \textit{$T_c$} is found. The rapid increase of the charge carrier density inferred from the suppression of the semiconducting behavior correlates with the rapid increase of \textit{$T_c$} in this region. At pressures above the value at which the maximum in \textit{$T_c$} occurs, there is a gradual decrease of \textit{$T_c$} and further suppression of the semiconducting behavior with pressure, both of which are reversible.

\begin{figure*}[t]
    \begin{minipage}[l]{1.0\columnwidth}
        \centering
        \includegraphics[scale=0.38, trim= 2.5cm 1.2cm 0cm 2cm, clip=true]{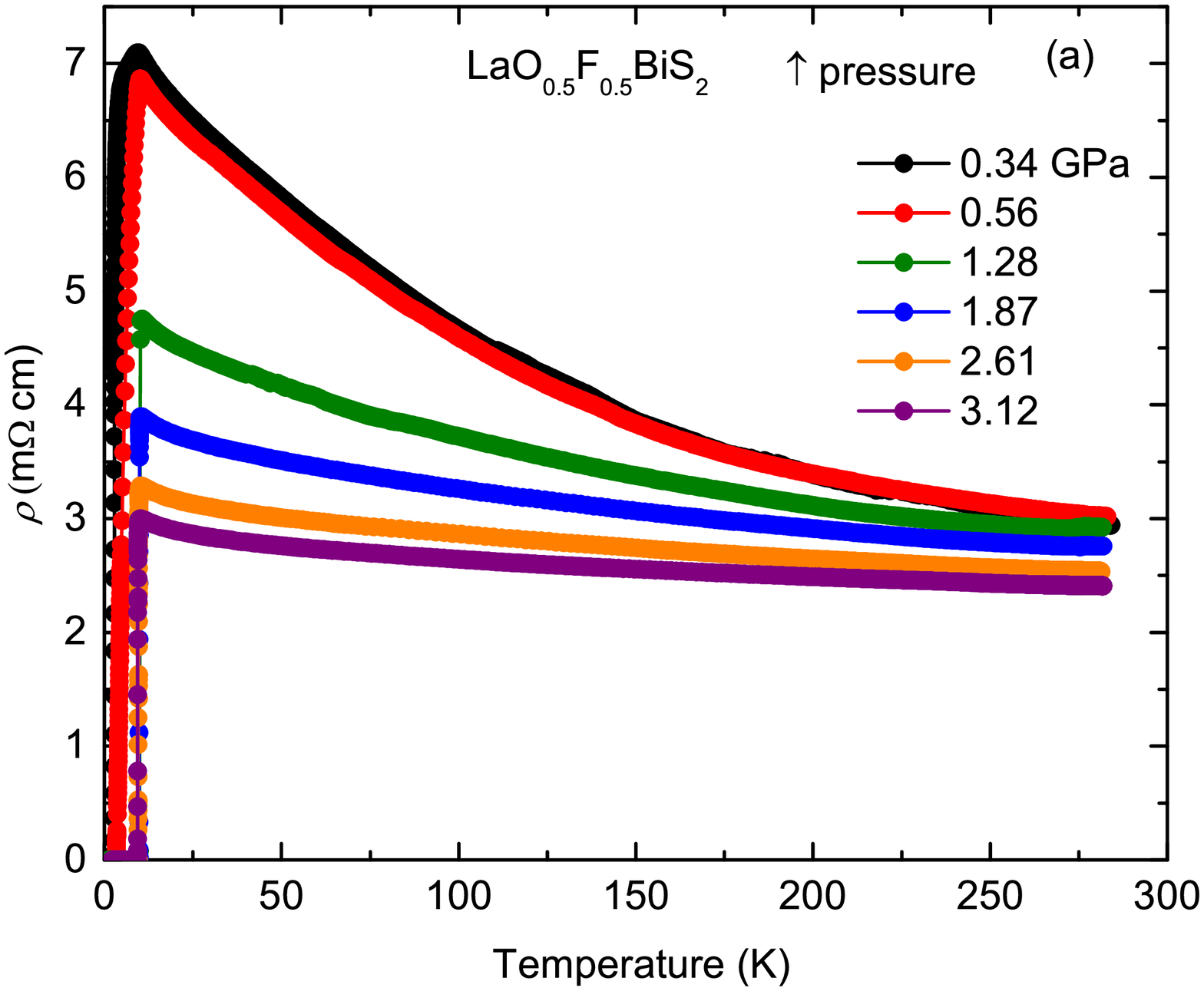}
    \end{minipage}
    \hfill{}
    \begin{minipage}[r]{1.0\columnwidth}
        \centering
        \includegraphics[scale=0.38, trim= 2.5cm 1.2cm 0cm 2cm, clip=true]{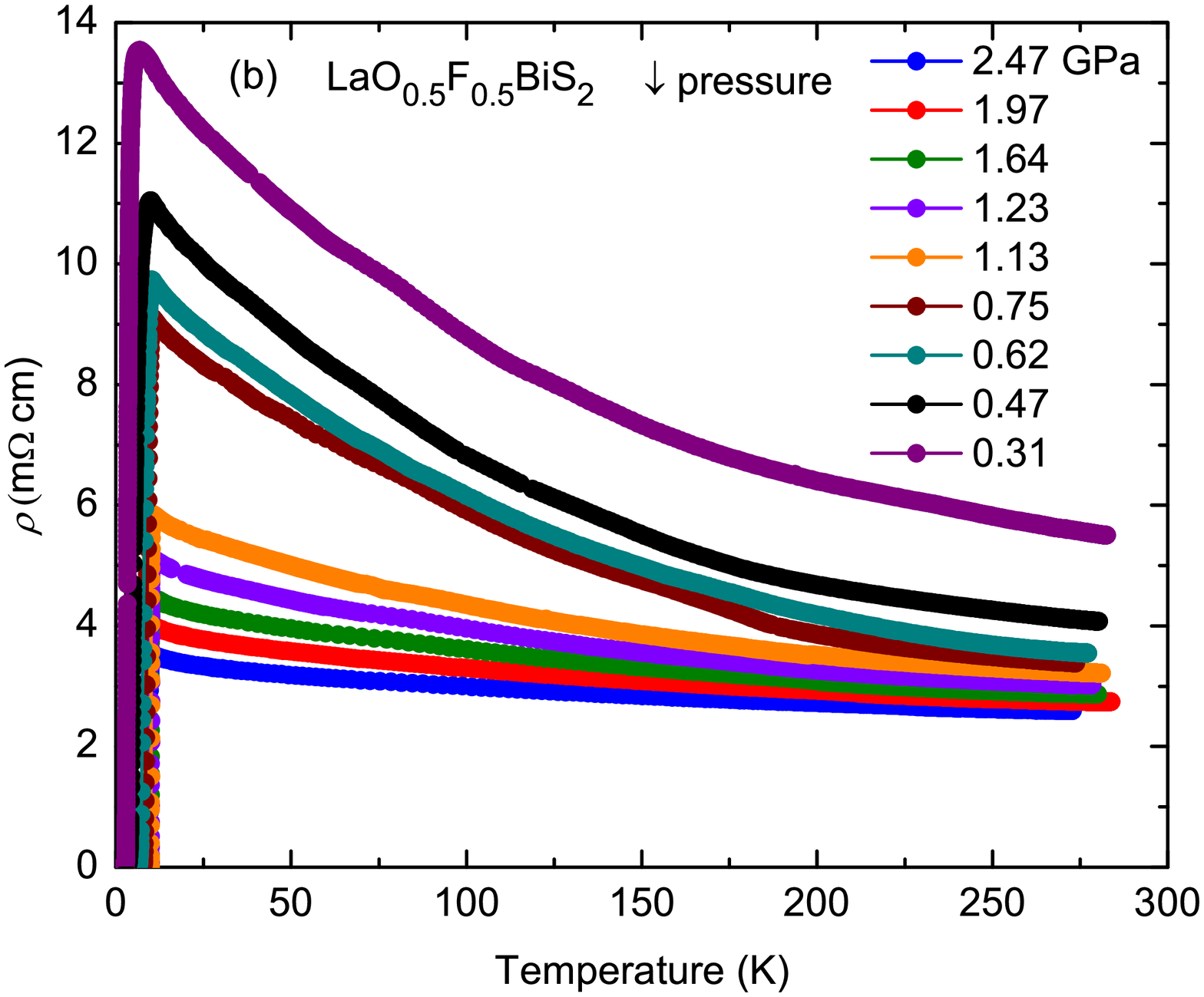}
            \end{minipage}
       \begin{minipage}[l]{1.0\columnwidth}
        \centering
        \includegraphics[scale=0.38, trim= 2.5cm 1.2cm 0cm 1cm, clip=true]{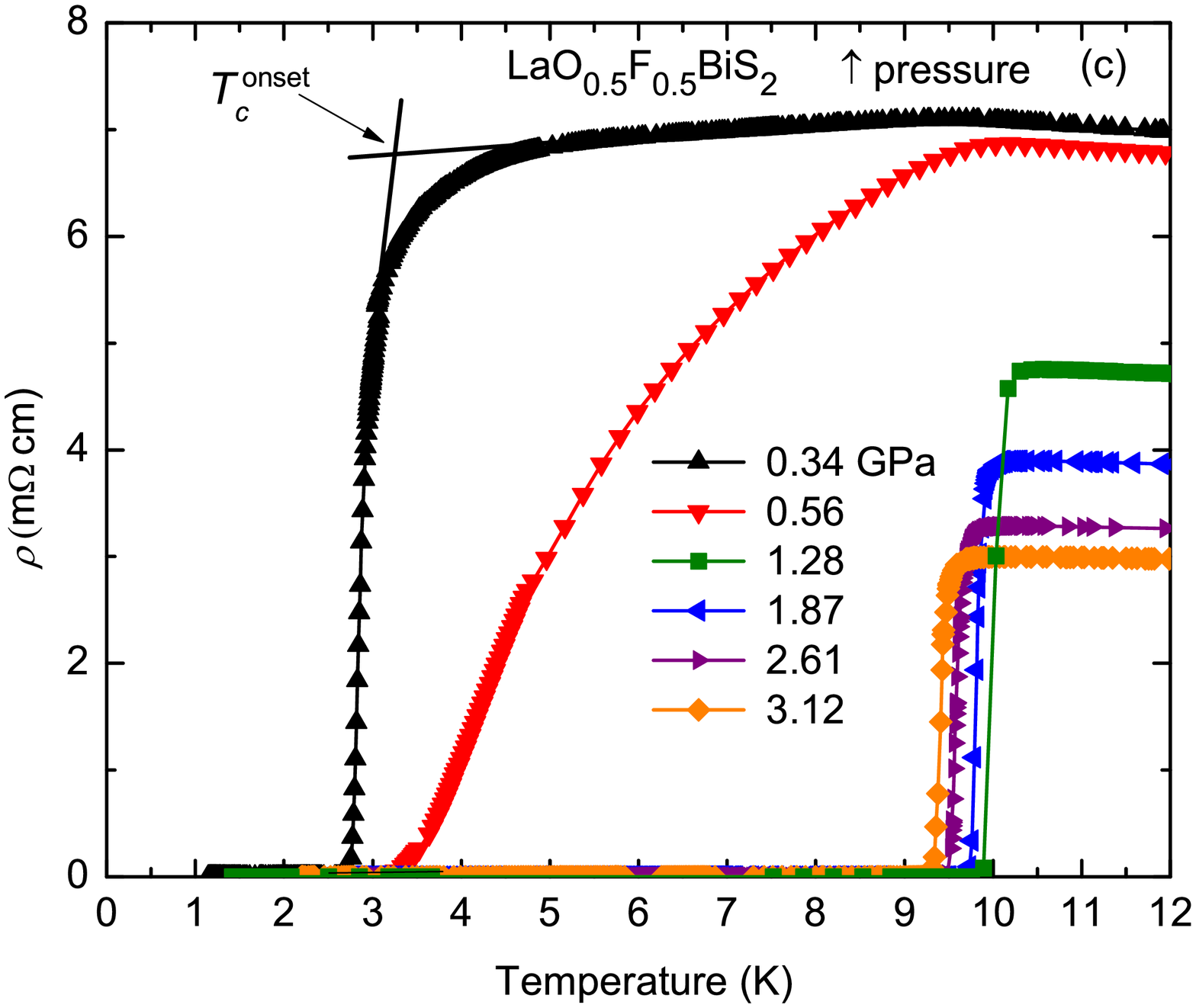}
      \end{minipage}
    \hfill{}
    \begin{minipage}[r]{1.0\columnwidth}
        \centering
        \includegraphics[scale=0.38, trim= 2.5cm 1.2cm 0cm 1cm, clip=true]{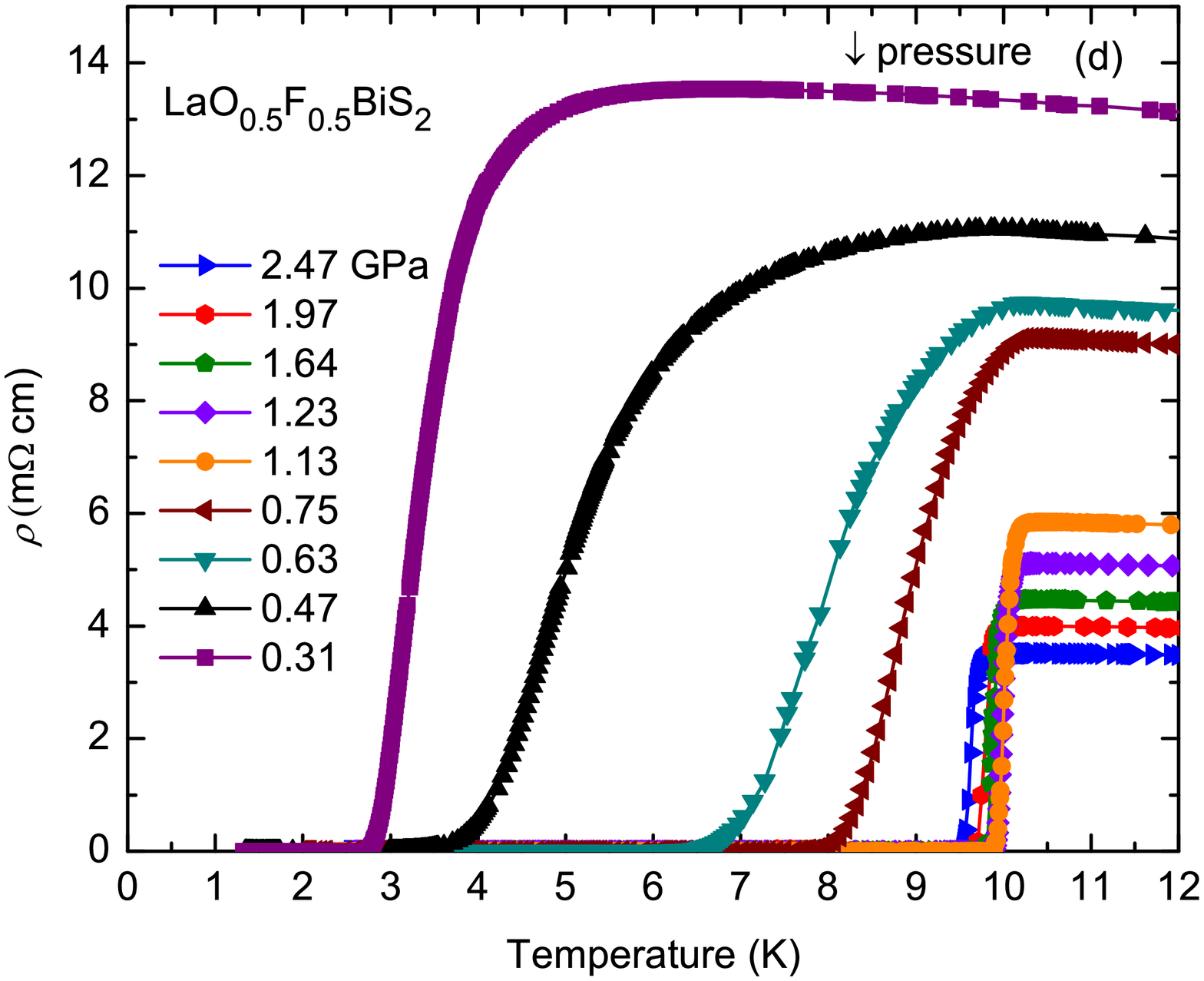}
      \end{minipage}
        \begin{minipage}[l]{2.0\columnwidth}
        \caption{\label{Lanthanum} (Color online) (a) (b) Temperature dependence of electrical resistivity, $\rho$, for LaO$_{0.5}$F$_{0.5}$BiS$_{2}$ at various pressures upon (a) increasing and (b) decreasing pressure. The electrical resistivity $\rho(T)$ is suppressed with increasing pressure as seen from the flattening of the curves at higher pressure. (c) (d) Resistive superconducting transition curves for LaO$_{0.5}$F$_{0.5}$BiS$_{2}$ upon (c) increasing and (d) decreasing pressure.  \textit{$T_c$} increases from $\sim$ 3 K to a maximum of $\sim$ 10 K before gradually decreasing. }
         \end{minipage}
\end{figure*}

\section{EXPERIMENTAL DETAILS}

Polycrystalline samples of \textit{Ln}O$_{1-x}$F$_{x}$BiS$_{2}$ (\textit{Ln} = La, Ce) with $x$ = 0.5 were prepared by solid-state reaction using powders of La$_{2}$O$_{3}$ (99.9\%), LaF$_{3}$ (99.9\%), La$_{2}$S$_{3}$ (99.9\%), and Bi$_{2}$S$_{3}$ (99.9\%) for LaO$_{1- x}$F$_{x}$BiS$_{2}$, and powders of CeF$_{3}$ (99.9\%) and CeO$_{2}$ (99.9\%) for CeO$_{1- x}$F$_{x}$BiS$_{2}$. Bi$_{2}$S$_{3}$ precursor powder was prepared in an evacuated quartz tube by reacting Bi (99.99\%) and S (99.9\%) at 500$^{\circ}$C for 10 hours.  The  \textit{Ln}$_{2}$S$_{3}$ (\textit{Ln} = La, Ce) precursor powders were  prepared in an evacuated quartz tube by reacting chunks of La and Ce with S grains at 800$^{\circ}$C for 12 hours. The starting materials with nominal composition \textit{Ln}O$_{0.5}$F$_{0.5}$BiS$_{2}$ (\textit{Ln} = La, Ce) were  weighed, thoroughly mixed, pressed into pellets, sealed in evacuated quartz tubes, and annealed at 800$^{\circ}$C for 10 hours. The products were then ground, mixed for homogenization, pressed into pellets, and annealed again in evacuated quartz tubes at 800$^{\circ}$C for 10 hours. X-ray powder diffraction measurements were made using an X-ray diffractometer with a Cu K$_{\alpha}$ source to assess phase purity and to determine the crystal structure of the \textit{Ln}O$_{0.5}$F$_{0.5}$BiS$_{2}$ (\textit{Ln} = La, Ce) compounds. Lattice parameters for LaO$_{0.5}$F$_{0.5}$BiS$_{2}$ are \textit{a} = 4.0613 \AA\ and \textit{c} = 13.3157 \AA,\ while for CeO$_{0.5}$F$_{0.5}$BiS$_{2}$ the lattice parameters are \textit{a} = 4.0398 \AA\  and \textit{c} = 13.4513 \AA. \cite{Yazici}\\

\indent Measurements of $\rho(T)$ under applied pressure were performed up to  $\sim$ 3 GPa in a clamped piston cylinder pressure cell between $\sim$ 1 K  and 300 K in a pumped $^4$He dewar. A 1:1 by volume mixture of $n$-pentane and isoamyl alcohol was used to provide a quasi-hydrostatic pressure transmitting medium. A second set of electrical resistivity measurements were performed by releasing pressure from the pressurized cell down to atmospheric pressure. Annealed Pt leads were affixed to gold-sputtered contact surfaces on each sample with silver epoxy in a standard four-wire configuration.   The pressure dependent superconducting  \textit{$T_c$} of high purity Sn, measured inductively, was used as a manometer and calibrated against data from Ref.~\onlinecite{Smith69}.

\begin{figure*}[t]
    \begin{minipage}[l]{1.0\columnwidth}
        \centering
        \includegraphics[scale=0.38, trim= 2.5cm 1.2cm 0cm 2cm, clip=true]{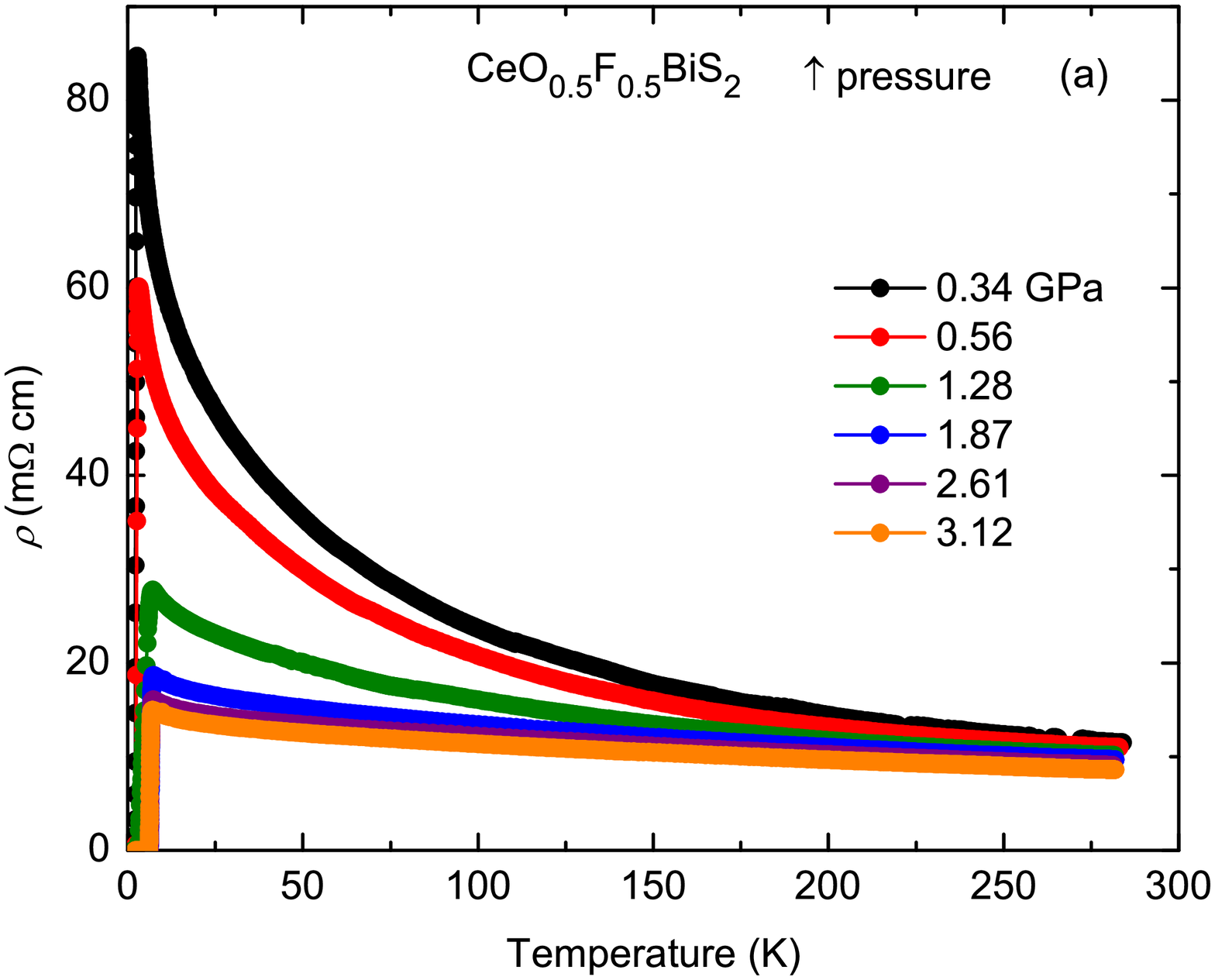}
    \end{minipage}
    \hfill{}
    \begin{minipage}[r]{1.0\columnwidth}
        \centering
        \includegraphics[scale=0.42, trim= 2.1cm 1.2cm 0cm 1.9cm, clip=true]{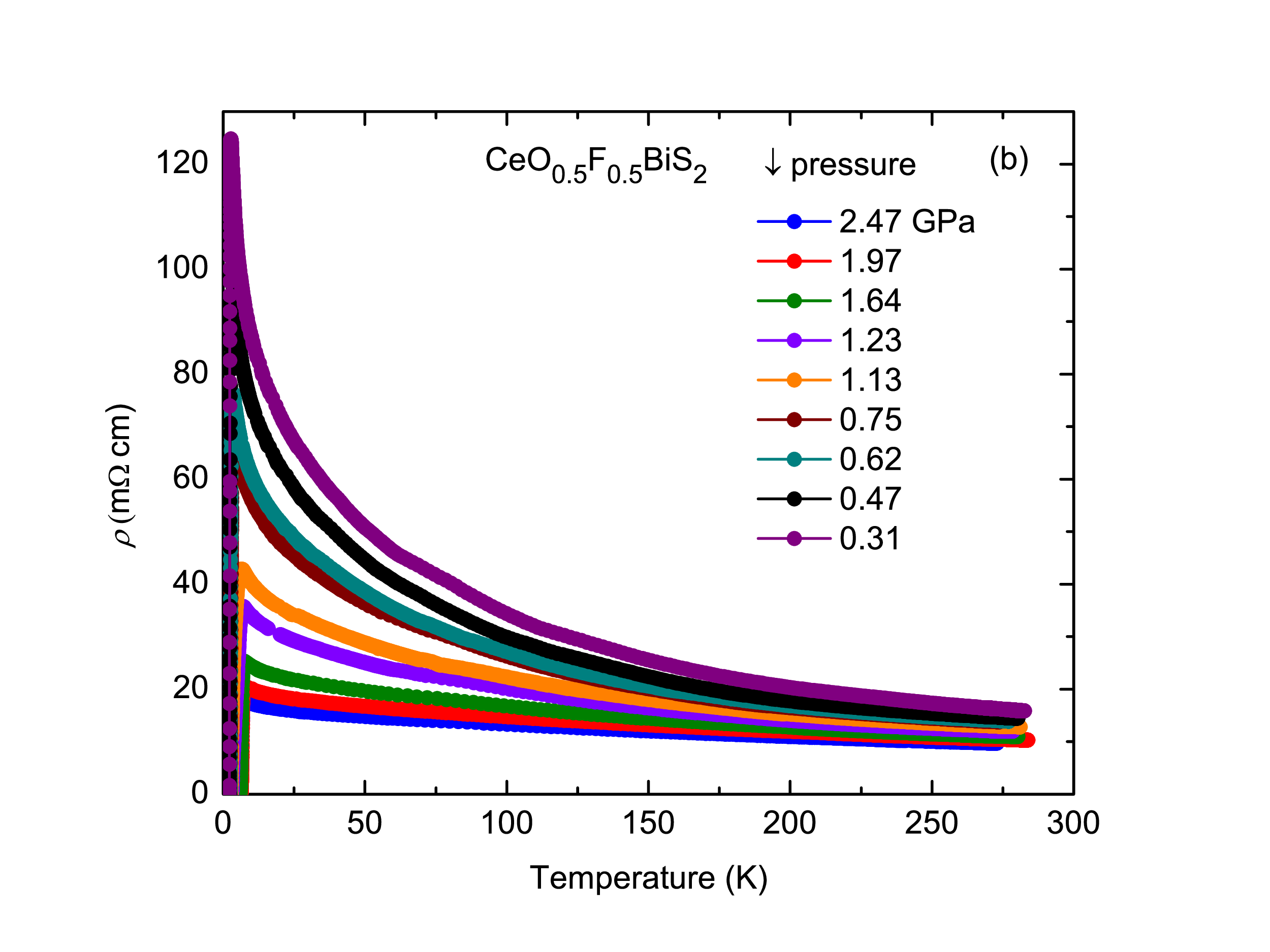}
            \end{minipage}
       \begin{minipage}[l]{1.0\columnwidth}
        \centering
        \includegraphics[scale=0.38, trim= 2.5cm 1.2cm 0cm 1cm, clip=true]{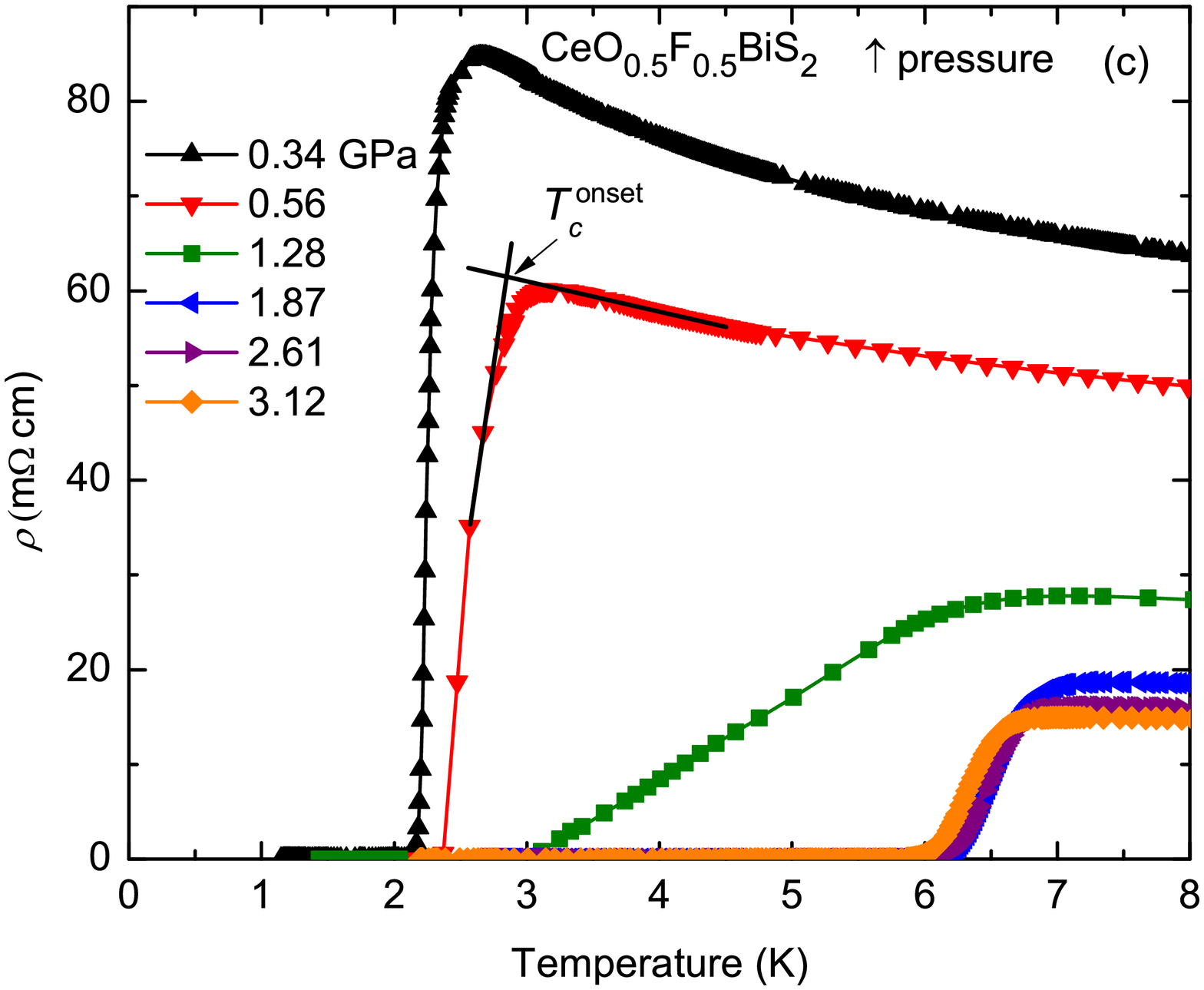}
      \end{minipage}
    \hfill{}
    \begin{minipage}[r]{1.0\columnwidth}
        \centering
        \includegraphics[scale=0.38, trim= 2.25cm 1.2cm 0cm 1cm, clip=true]{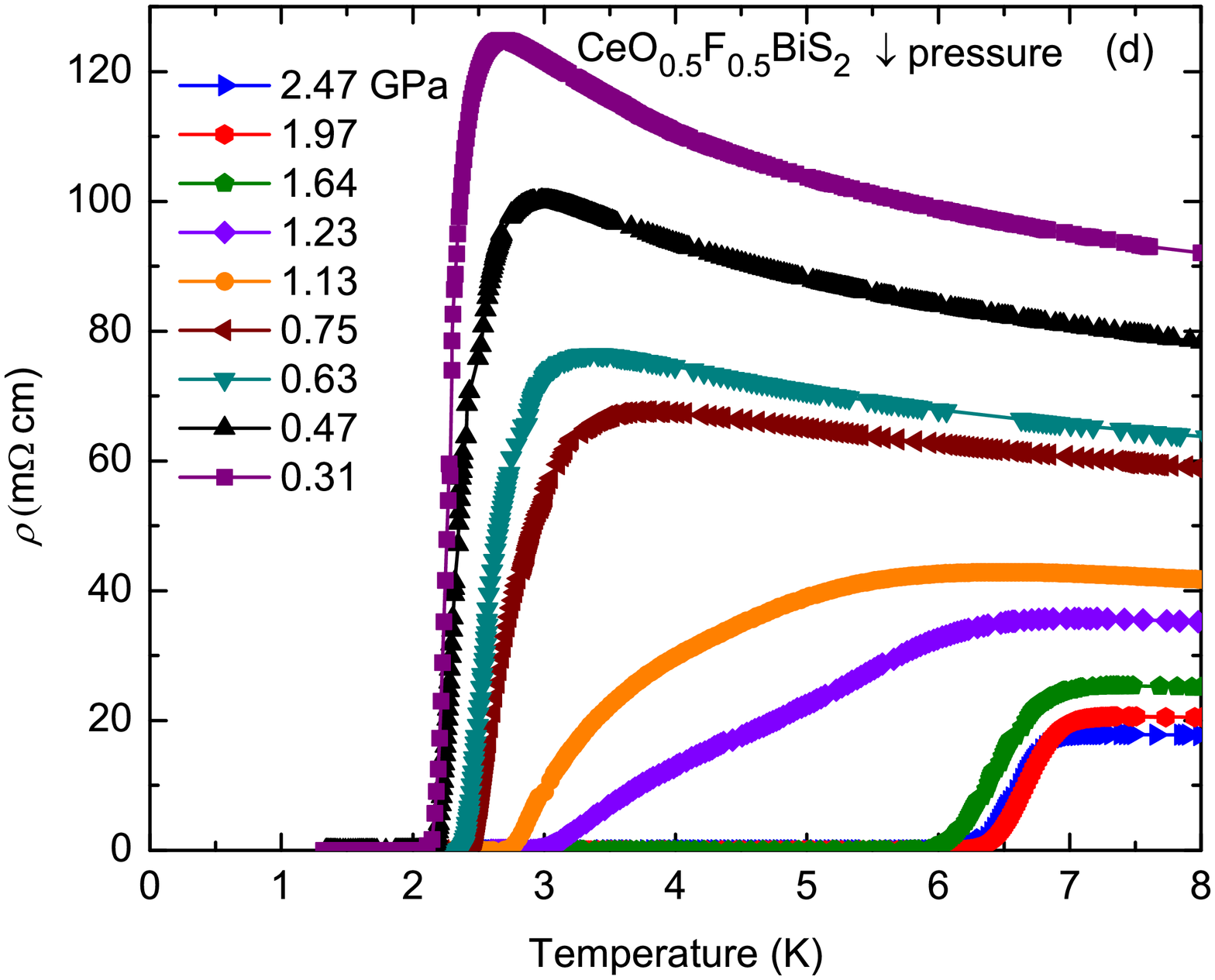}
      \end{minipage}
        \begin{minipage}[l]{2.0\columnwidth}
         \caption{\label{Cerium} (Color online) (a) (b) Temperature dependence of electrical resistivity, $\rho$, for CeO$_{0.5}$F$_{0.5}$BiS$_{2}$ at various pressures upon (a) increasing pressure and (b) decreasing pressure. At lower pressures, the compound exhibits semiconducting behavior. The semiconducting behavior is suppressed at higher pressure as seen from the flattening of the curves. (c) (d) Resistive superconducting transition curves for CeO$_{0.5}$F$_{0.5}$BiS$_{2}$ at various pressures upon (c) increasing and (d) decreasing pressure.  \textit{$T_c$} increases from $\sim$ 2.2 K to a maximum of $\sim$ 6.7 K and then gradually decreases. }
        \end{minipage}
\end{figure*}

\section{RESULTS AND DISCUSSION}
The temperature dependence of the electrical resistivity, $\rho$, below 300 K for LaO$_{0.5}$F$_{0.5}$BiS$_{2}$ at various pressures is displayed in Fig.~\ref{Lanthanum}. Figure~\ref{Lanthanum}(a) shows $\rho(T)$ upon increasing pressure to 3.1 GPa, while Fig.~\ref{Lanthanum}(b) gives $\rho(T)$ upon decreasing pressure back down to 0.31 GPa. The temperature dependence of $\rho(T)$ at lower pressures exhibits semiconducting behavior. The semiconducting behavior is suppressed with increasing pressure as seen from the nearly constant $\rho(T)$ curves above 2 GPa. A comparison of Fig.~\ref{Lanthanum}(a) with  Fig.~\ref{Lanthanum}(b) shows that the suppression of $\rho(T)$ is continuous and reversible over the full range 0.3 - 3.1 GPa. At pressures above 2 GPa where suppression is greatest, the values of $\rho(T)$ are comparable in Fig.~\ref{Lanthanum}(a) and Fig.~\ref{Lanthanum}(b) and reversible with pressure.  However, whereas the suppression of $\rho(T)$ is reversible with pressure, the magnitude of $\rho(T)$ exhibits hysteretic behavior with pressure.  At lower pressures, measurements of $\rho(T)$ made during release of pressure yield higher values than $\rho(T)$ measurements performed upon increasing pressure. The difference between $\rho(T)$ measurements made during increasing pressure and those made during decreasing pressure are largest at lower pressures where the rate of suppression of semiconducting behavior is largest. After a release in pressure to the lowest value of 0.31 GPa, the maximum value of $\rho$ is nearly 14 m$\Omega$ cm. This is a factor of 2 larger than the corresponding value along the increasing pressure path at 0.34 GPa.\\

\indent Superconducting transitions at low temperature were measured upon increasing and then releasing pressure as shown in Fig.~\ref{Lanthanum}(c) and Fig.~\ref{Lanthanum}(d), respectively. There is a striking similarity in the qualitative behavior and evolution of the transitions in both plots. For pressures in the range 0.5 GPa to 1.0 GPa, the superconducting transitions broaden significantly. For higher pressures above 1.0 GPa, the transition curves begin to sharpen again at approximately 10 K. It is in this higher pressure region where \textit{$T_c$} passes through a maximum of 10.1 K at $\sim$ 1 GPa and then gradually decreases as pressure increases. The evolution of both the value of the superconducting critical temperature \textit{$T_c$} and the superconducting transition width $\Delta$\textit{$T_c$} defined by the procedure described in the text, were reversible with respect to both increasing and decreasing pressure. \textit{$T_c$} was defined as the temperature at which $\rho$ falls to 50\% of its value at the temperature of the onset of superconductivity, \textit{$T^{onset}_c$}, with \textit{$T^{onset}_c$} determined as illustrated in Fig.~\ref{Lanthanum}(c). The temperature where the resistivity vanishes, \textit{$T_0$}, was determined in a similar fashion as \textit{$T^{onset}_c$} using a linear extrapolation of the resistive superconducting curve to $\rho$ = 0. In determining \textit{$T_c$} for the broader transitions, we used the same criteria as for the sharper transitions; however, we make note of the less definitive \textit{$T_c$}  for these broader transitions.\\

\begin{figure}[t]
\includegraphics[scale=0.43, trim= 2.4cm .9cm 0cm 2cm, clip=true]{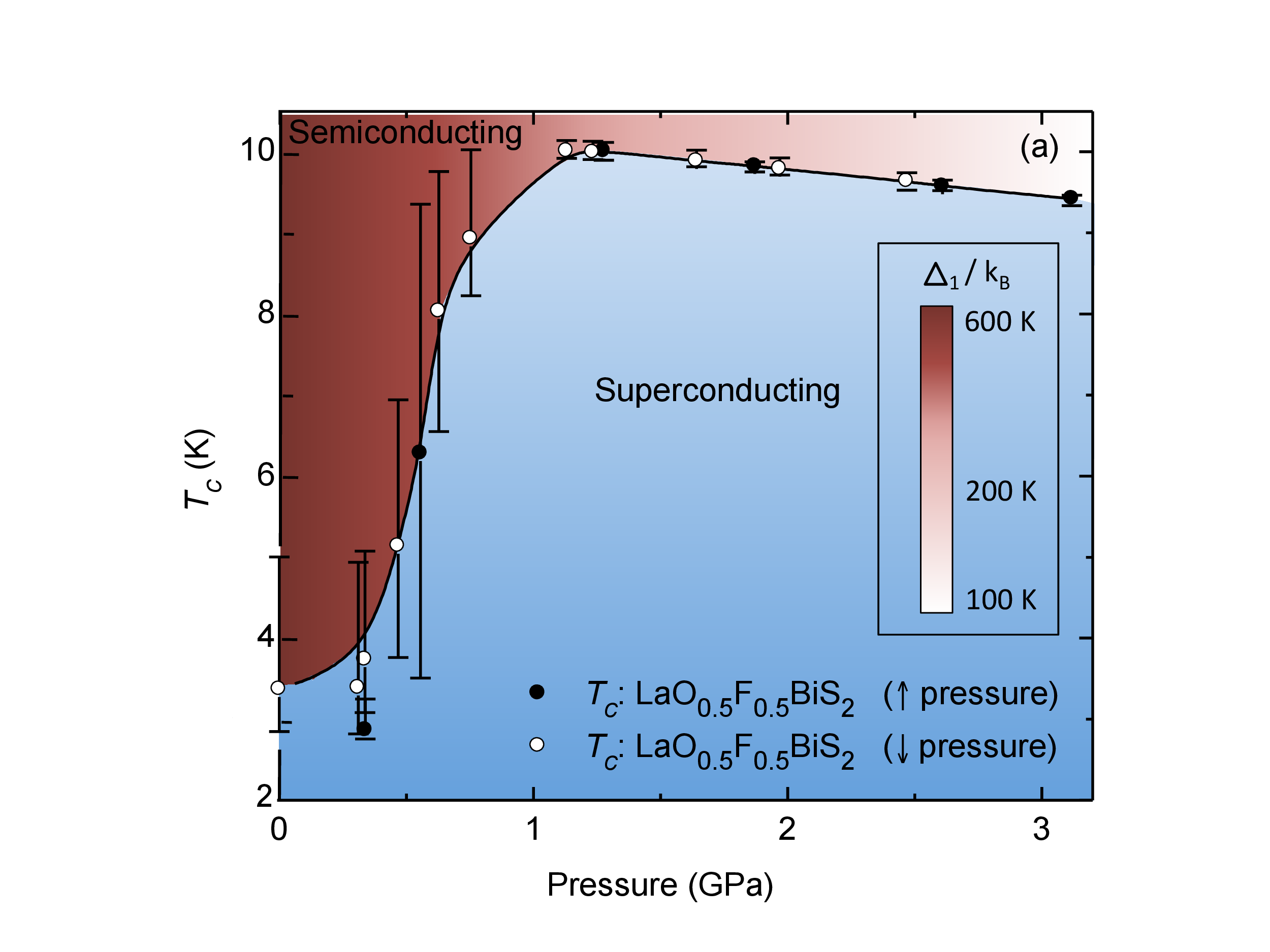}
\includegraphics[scale=0.43, trim= 2.4cm .9cm 0cm 1cm, clip=true]{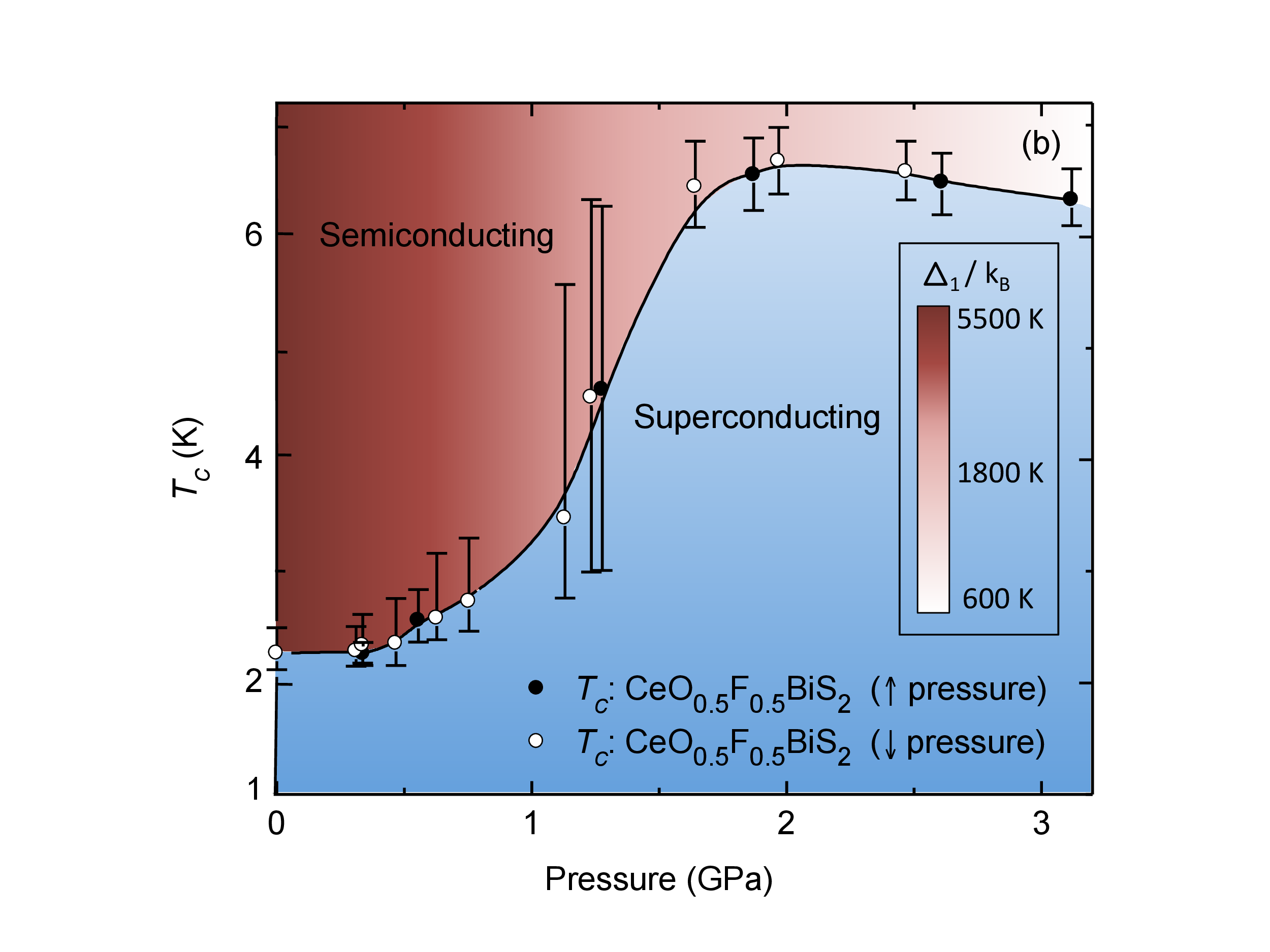}
\caption{\label{Tc Combined Spline}  (Color online) Phase diagrams for (a) LaO$_{0.5}$F$_{0.5}$BiS$_{2}$ and (b) CeO$_{0.5}$F$_{0.5}$BiS$_{2}$ under pressure. \textit{$T_c$} was defined as the temperature at which the electrical resistivity, $\rho$, falls to 50\% of its value at the temperature of the onset of superconductivity, \textit{$T^{onset}_c$}. Vertical bars indicate the superconducting transition width $\Delta$\textit{$T_c$} and vertical bar caps indicate \textit{$T_c^{onset}$} (upper) and \textit{$T_0$} (lower). Filled symbols denote measurements performed upon increasing pressure, while open symbols represent measurements made upon decreasing pressure. The solid black curves are guides to the eye. The change in color in the semiconducting normal state region represents the suppression of the semiconducting behavior with pressure as manifested in the decrease of the larger energy gap $\Delta_1$ whose values are indicated in the legend.}
\end{figure}

\indent Measurements performed on CeO$_{0.5}$F$_{0.5}$BiS$_{2}$ reveal remarkably similar behavior to the LaO$_{0.5}$F$_{0.5}$BiS$_{2}$ results. As shown in Fig.~\ref{Cerium}, the qualitative behavior of the results are reversible upon application and release of pressure. Measurements of $\rho(T)$ show semiconducting behavior which is suppressed at higher pressures. As pressure is released, the semiconducting behavior is recovered. The measured values of $\rho(T)$ are higher along the reversed path during a release of pressure. The discrepancy between $\rho(T)$ measurements made during increasing pressure and those made during decreasing pressure are largest at lower pressures where the rate of suppression of the semiconducting behavior is largest. After releasing the pressure to the lowest value of 0.31 GPa, $\rho$ is nearly 130 m$\Omega$ cm; this is a factor of 1.5 larger than the corresponding value along the increasing pressure path at 0.34 GPa.\\

\begin{figure}[t]
\includegraphics[scale=0.38, trim= 2.5cm 1.2cm 0cm 2cm, clip=true]{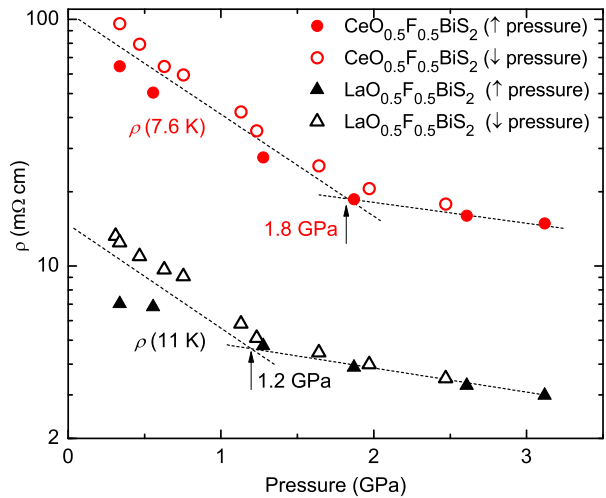}
\caption{\label{Resistivity}  (Color online) Electrical resistivity, $\rho$, in the normal state just above the superconducting onset temperature, \textit{$T_c^{onset}$}.  Electrical resistivity values for CeO$_{0.5}$F$_{0.5}$BiS$_{2}$ were taken
at \textit{T} = 7.6 K, while $\rho$ values for LaO$_{0.5}$F$_{0.5}$BiS$_{2}$ were taken at \textit{T} = 11 K. Filled (open) symbols represent measurements upon increasing (decreasing) pressure. Dotted lines reflect the slopes (suppression rates), and arrows point to changing slopes at $\sim$ 1.2 GPa and  $\sim$ 1.8 GPa in LaO$_{0.5}$F$_{0.5}$BiS$_{2}$ and CeO$_{0.5}$F$_{0.5}$BiS$_{2}$, respectively. The break in slope occurs at a pressure near that at which \textit{$T_c$} reaches a maximum for both LaO$_{0.5}$F$_{0.5}$BiS$_{2}$ and CeO$_{0.5}$F$_{0.5}$BiS$_{2}$.}
\end{figure}

Superconducting transitions at low temperature were measured for CeO$_{0.5}$F$_{0.5}$BiS$_{2}$ while increasing and then decreasing pressure. The trend and character of the transitions are reversible upon application and subsequent release of pressure as seen from a comparison of Fig.~\ref{Cerium}(c) and Fig.~\ref{Cerium}(d). Similar to the evolution of \textit{$T_c$} in LaO$_{0.5}$F$_{0.5}$BiS$_{2}$, sharp superconducting transitions are observed at low pressures up to approximately 0.5 GPa before they begin to broaden. From both the increasing and decreasing pressure plots (Fig.~\ref{Cerium}(c) and Fig.~\ref{Cerium}(d), respectively), the transitions begin to broaden significantly up to pressures of approximately 1.5 GPa. At pressures above 1.5 GPa, the superconducting transitions become sharp again. It is in this pressure region where \textit{$T_c$} passes through a maximum of 6.7 K at $\sim$ 2 GPa and then decreases gradually at higher pressures.\\

Figure~\ref{Tc Combined Spline} summarizes the results for the superconducting phase diagram, \textit{$T_c(P)$}, for both the LaO$_{0.5}$F$_{0.5}$BiS$_{2}$ and CeO$_{0.5}$F$_{0.5}$BiS$_{2}$ compounds. The measurements were performed first by increasing the pressure monotonically in six steps (filled symbols) up to 3.1 GPa, followed by a monotonic decrease in pressure in nine steps (open symbols) back down to 0.31 GPa. The phase diagram indicates \textit{$T_c(P)$} is highly reversible for both compounds; negligible pressure hysteresis is observed even in the regions where the resistive transition broadens significantly.\\

The phase diagram in Fig.~\ref{Tc Combined Spline} shows \textit{$T_c$} maxima for both the LaO$_{0.5}$F$_{0.5}$BiS$_{2}$ and CeO$_{0.5}$F$_{0.5}$BiS$_{2}$ compounds. In the case of LaO$_{0.5}$F$_{0.5}$BiS$_{2}$, \textit{$T_c$} initially increases with pressure up to a maximum of 10.1 K at 1.1 GPa. \textit{$T_c$} then gradually decreases with a slope of -0.30 K GPa$^{-1}$. This  \textit{$T_c$} maximum is also preceded by a reversible broadening of the superconducting transition in the pressure range 0.5 - 1.0 GPa, represented as elongated vertical bars in Fig.~\ref{Tc Combined Spline}.  In the normal state (above the \textit{$T_c(P)$} curve), the semiconducting behavior of LaO$_{0.5}$F$_{0.5}$BiS$_{2}$ is continuously suppressed with pressure as manifested in a larger energy gap $\Delta_1$ that decreases with pressure, the values of which are indicated in the legend.  The determination of the energy gaps $\Delta_1$ and $\Delta_2$  from the $\rho(T,P)$ data are discussed below.  This maximum in \textit{$T_c(P)$} at 1.1 GPa occurs in the vicinity of a change of slope in the normal state $\log(\rho)$ vs. $P$ curve, measured at 11 K, as shown in Fig.~\ref{Resistivity}.\\

\begin{figure}[t]
\includegraphics[scale=0.38, trim= 2.3cm 1.2cm 0cm 2.0cm, clip=true]{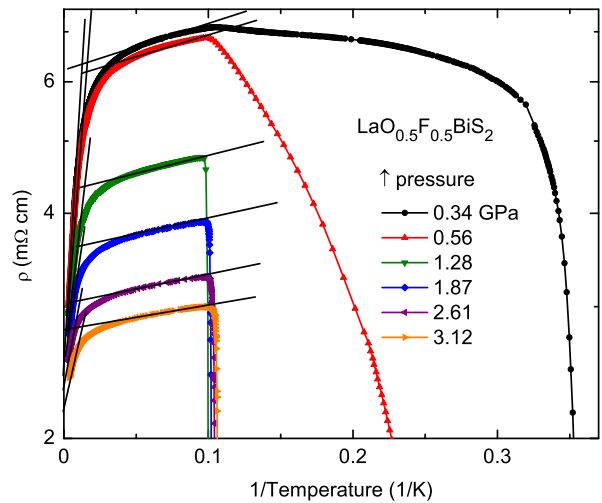}
\caption{\label{Lanthanum_Gap}  (Color online) $\log(\rho)$  vs. 1/\textit{T} up to 3.12 GPa for LaO$_{0.5}$F$_{0.5}$BiS$_{2}$. The solid lines represent linear fits of Eq. \eqref{equation 1} from which the high and low temperature gaps $\Delta_1$ and $\Delta_2$, respectively, were determined.}
\end{figure}

In the case of CeO$_{0.5}$F$_{0.5}$BiS$_{2}$, \textit{$T_c$} initially increases to a maximum of 6.7 K at $\sim$ 2 GPa and then decreases slowly with a slope of -0.30 K GPa$^{-1}$ at higher pressures. Leading up to this maximum is a reversible broadening of the superconducting transition in the region 0.5 - 1.5 GPa depicted by the lengthened vertical bars in Fig.~\ref{Tc Combined Spline}. In the normal state (above the \textit{$T_c(P)$} curve), the semiconducting behavior of CeO$_{0.5}$F$_{0.5}$BiS$_{2}$ is also continuously suppressed with pressure as manifested in a larger energy gap $\Delta_1$ that decreases with pressure, the values of which are indicated in the legend. The maximum in \textit{$T_c(P)$} at 2 GPa also occurs in the vicinity of a slope change in $\log(\rho)$ vs. $P$, measured at 7.6 K, as shown in Fig.~\ref{Resistivity}.\\

The width of the superconducting transitions in the broadening region, represented by the vertical bars in Fig.~\ref{Tc Combined Spline}, is $\Delta T_c \sim$ 4 - 6 K.  Pressure gradients in the piston-cylinder cell were estimated from the error in pressure to be of the order $\Delta P$ $\sim$ $\pm$ 0.05 GPa where the error in pressure was determined from the width of the superconducting transition of the Sn manometer.  It is possible to relate $\Delta T_c$ and $\Delta P$ through the slope of $T_c(P)$ in Fig.~\ref{Tc Combined Spline} so that $\Delta T_c \simeq \left(dT_c(P)/dP\right)\Delta P$.  Even though $\Delta P$ is small and constant for pressures measured as part of this study, $\Delta T_c$ can be large when $dT_c(P)/dP$ is large (\textit{i.e.}, in the pressure region where broadened transitions are observed).  Rough estimates of $\Delta T_c$ were made using $\Delta P$ = 0.1 GPa and slopes of 18 K/GPa and 11 K/GPa for LaO$_{0.5}$F$_{0.5}$BiS$_{2}$ and CeO$_{0.5}$F$_{0.5}$BiS$_{2}$, respectively.  These calculations yield values of $\Delta T_c$ = 1.8 K and 1.1 K for LaO$_{0.5}$F$_{0.5}$BiS$_{2}$ and CeO$_{0.5}$F$_{0.5}$BiS$_{2}$, respectively, which are of the correct order of magnitude.  The size of the vertical bars characterizing $\Delta T_c$ also appear to qualitatively track with the local slope of $T_c(P)$ for most pressures in Fig.~\ref{Tc Combined Spline}.\\

\begin{figure*}[t]
    \begin{minipage}[l]{1.0\columnwidth}
        \centering
        \includegraphics[scale=0.38, trim= 1.7cm 1.2cm 0cm 2cm, clip=true]{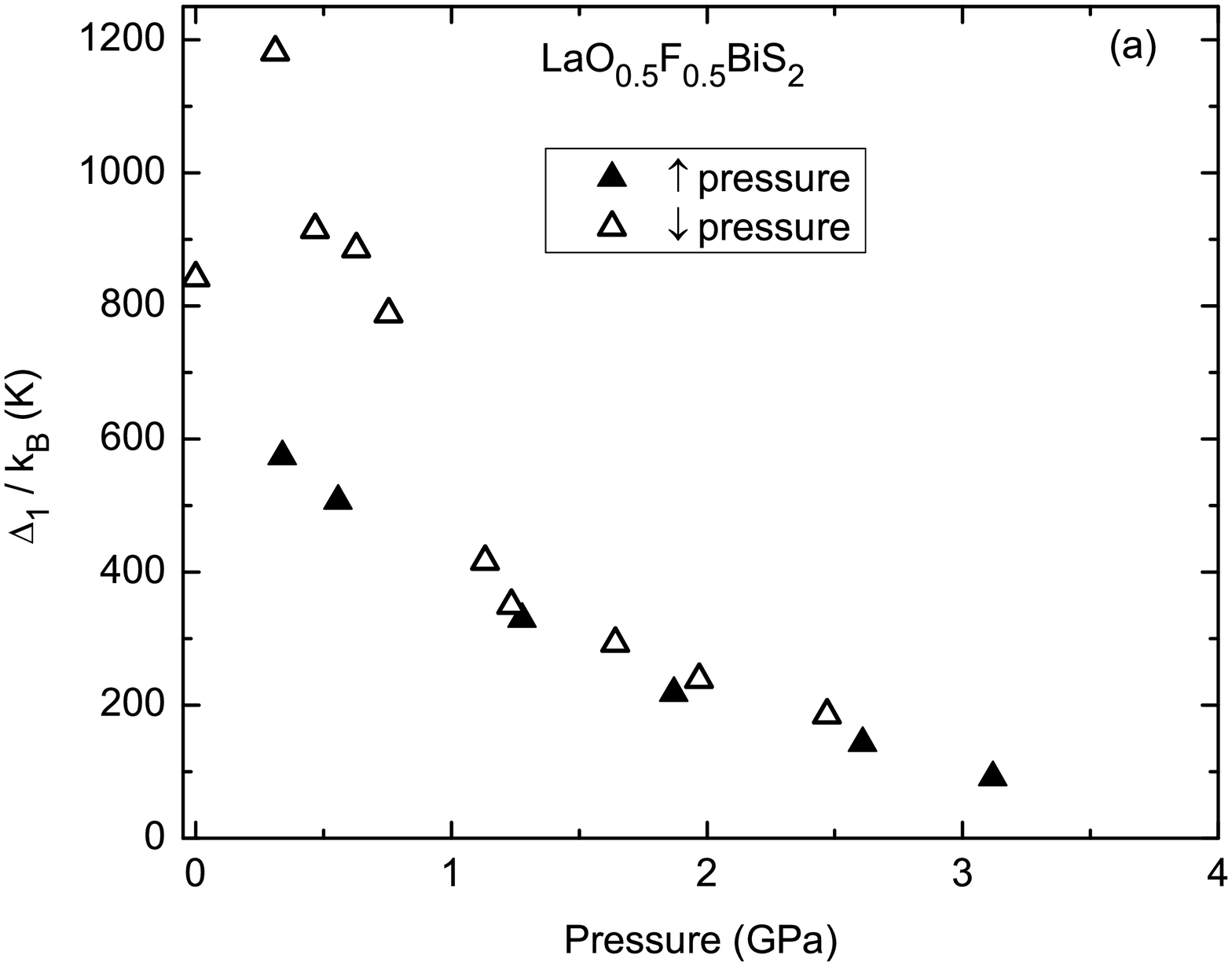}
    \end{minipage}
    \hfill{}
    \begin{minipage}[r]{1.0\columnwidth}
        \centering
        \includegraphics[scale=0.38, trim= 2.0cm 1.2cm 0cm 2cm, clip=true]{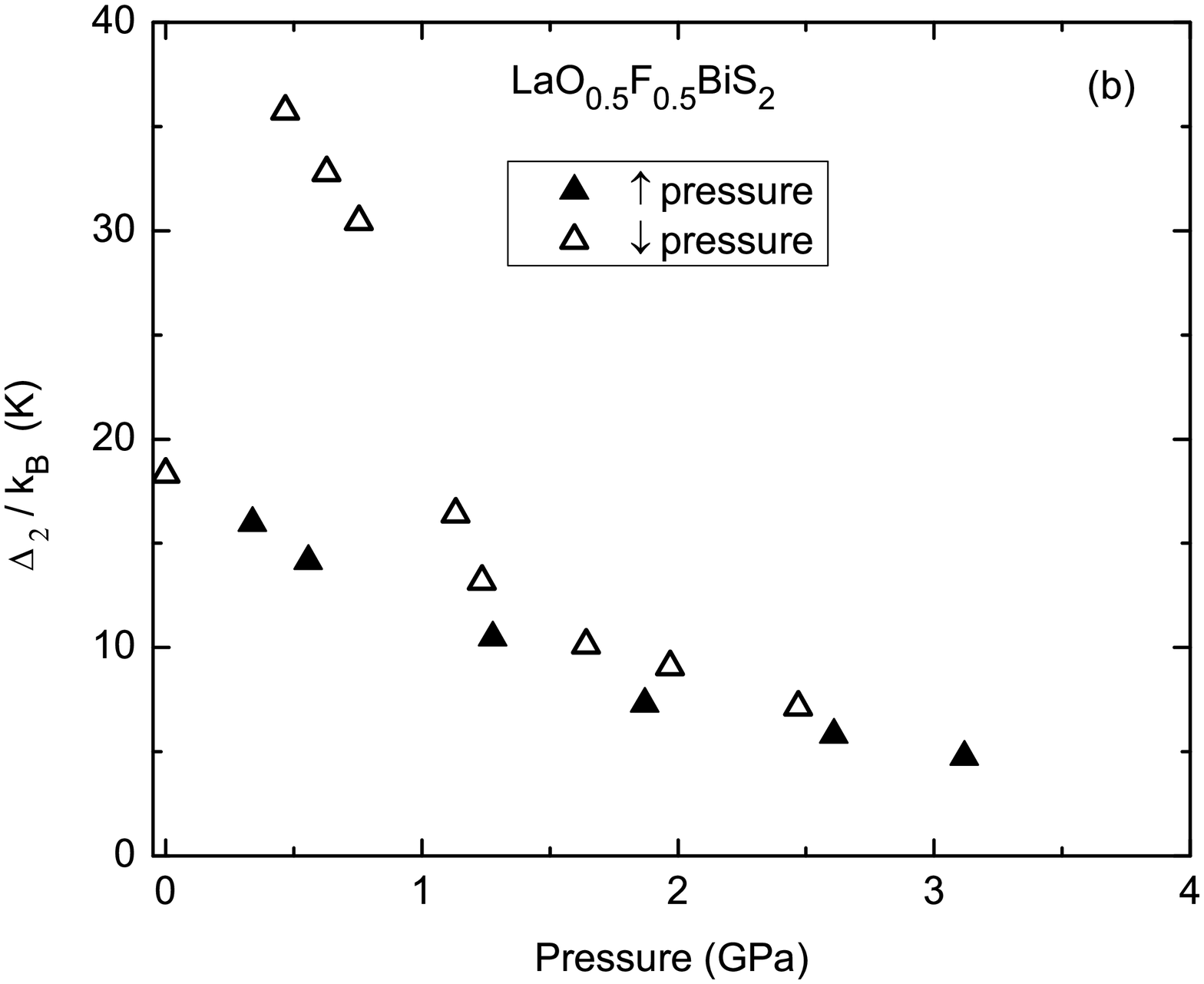}
            \end{minipage}
       \begin{minipage}[l]{1.0\columnwidth}
        \centering
        \includegraphics[scale=0.38, trim= 1.5cm 1.2cm 0cm 1cm, clip=true]{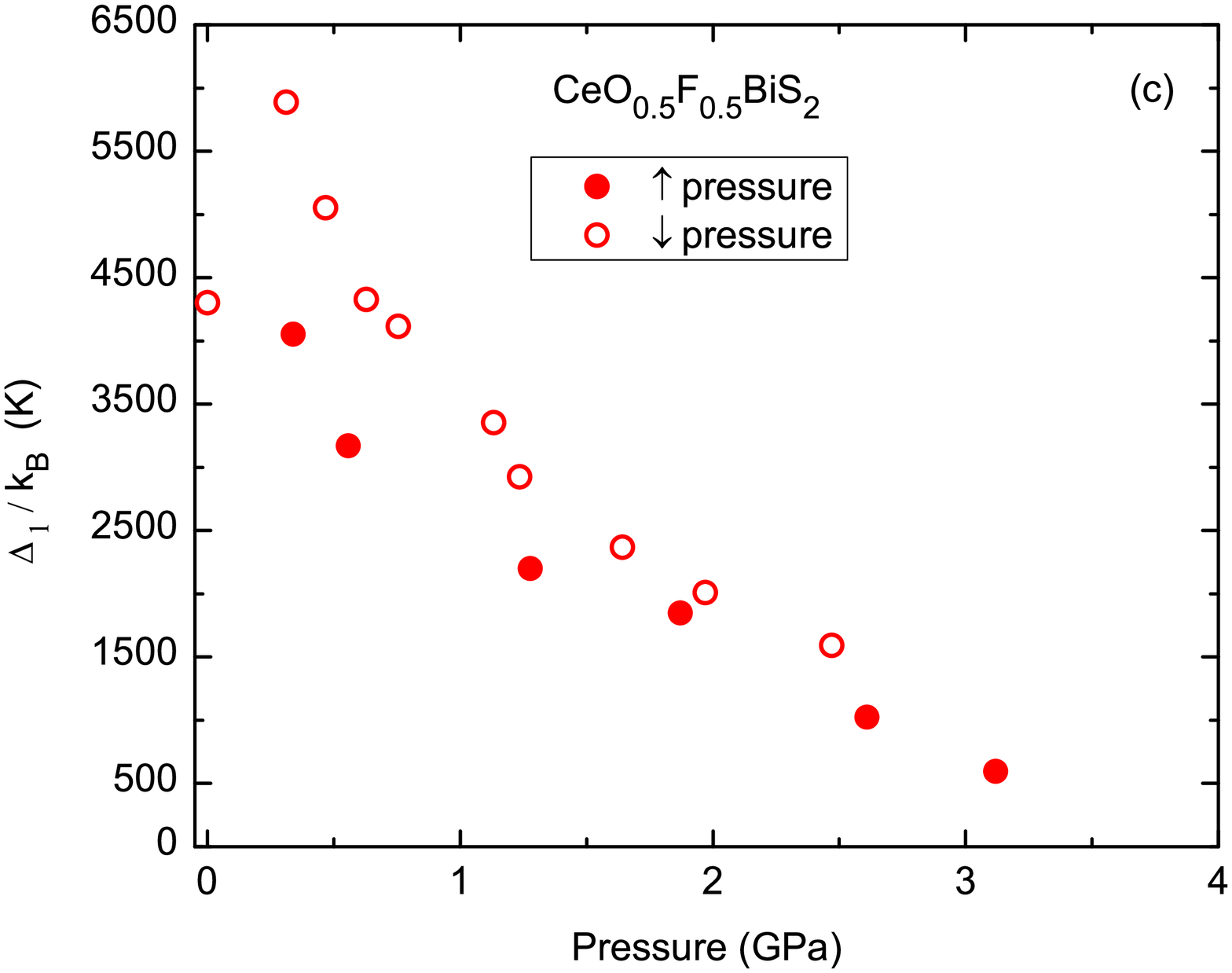}
      \end{minipage}
    \hfill{}
    \begin{minipage}[r]{1.0\columnwidth}
        \centering
        \includegraphics[scale=0.38, trim= 1.9cm 1.2cm 0cm 1cm, clip=true]{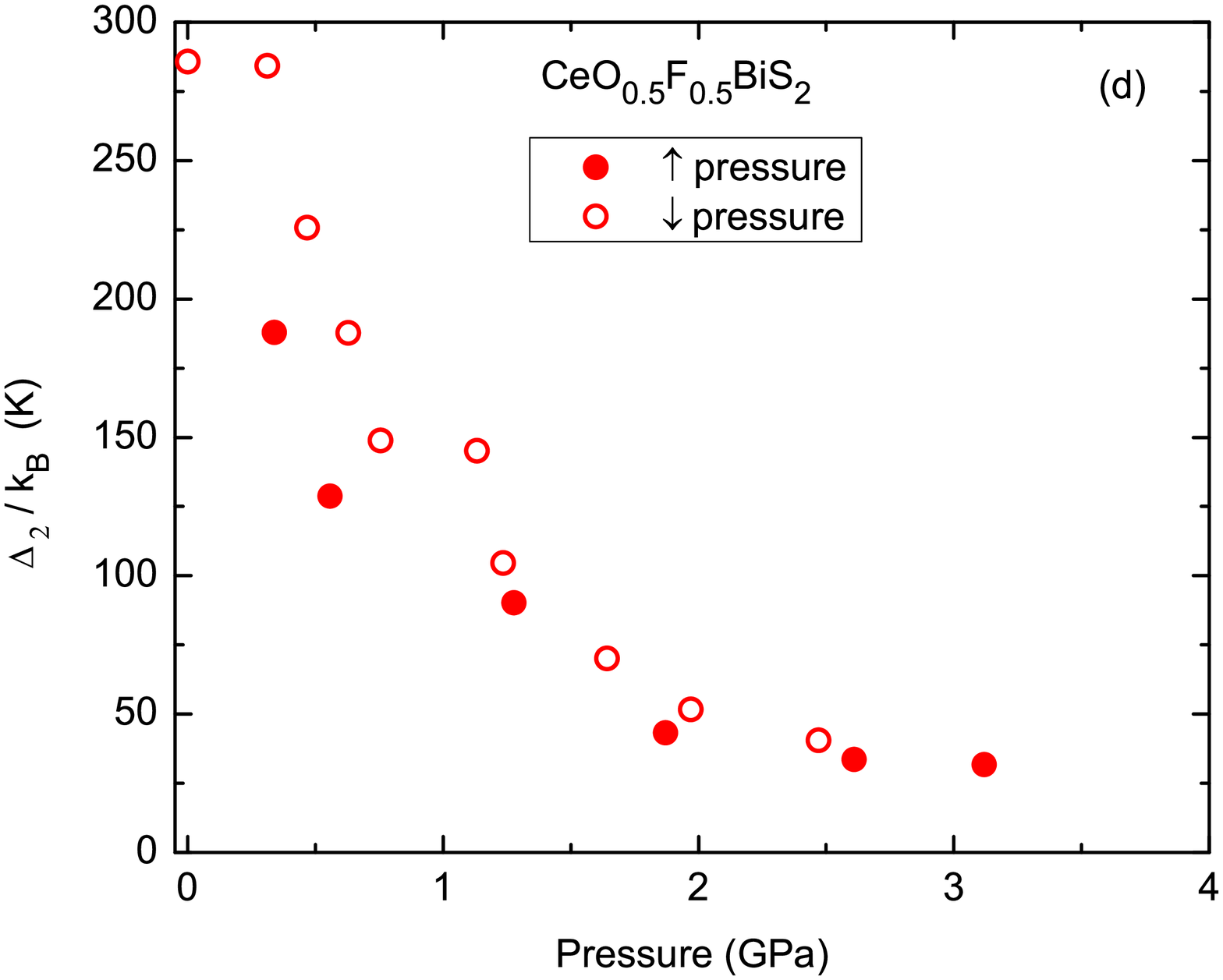}
      \end{minipage}
        \begin{minipage}[l]{2.0\columnwidth}
         \caption{\label{Gap_Pressure} (Color online) (a) (b)  Fitted energy gaps $\Delta_1$ and $\Delta_2$ plotted as a function of pressure for LaO$_{0.5}$F$_{0.5}$BiS$_{2}$ (c) (d) fitted energy gaps $\Delta_1$ and $\Delta_2$ plotted as a function of pressure for CeO$_{0.5}$F$_{0.5}$BiS$_{2}$. Gaps were fitted from the data shown in Fig.~\ref{Lanthanum_Gap}. Filled (open) symbols were obtained upon increasing (decreasing) pressure cycles. }
        \end{minipage}
\end{figure*}

\indent Kotegawa \textit{et al}.\cite{Kotegawa} previously reported the pressure dependence of \textit{$T_c$} for LaO$_{0.5}$F$_{0.5}$BiS$_{2}$ samples synthesized under high pressure, which apparently exhibit only the high  \textit{$T_c$}  phase uncovered in the present study. In their experiments, it was found that \textit{$T_c$} exhibits a maximum of 10.6 K at $\sim$1 GPa and then gradually decreases with a slope of -0.40 K/GPa (compared to -0.30 K/GPa in this study) at pressures above 1 GPa. The low \textit{$T_c$}  phase and broadened superconducting transitions bridging the low \textit{$T_c$} and high \textit{$T_c$} phases, however, are not present in their \textit{$T_c(P)$} phase diagram. The presence of only the high \textit{$T_c$} phase at ambient pressure in the study by Kotegawa \textit{et al}.\cite{Kotegawa} suggests that synthesis of the LaO$_{0.5}$F$_{0.5}$BiS$_{2}$ samples under high pressure has already induced the high \textit{$T_c$} superconducting phase.\\

From the plot of $\log(\rho)$ vs. $P$ at low temperature displayed in Fig.~\ref{Resistivity}, there is a noticeable change in the magnitude of the suppression rate, d$\log(\rho)$/d$P$, for both the LaO$_{0.5}$F$_{0.5}$BiS$_{2}$ and CeO$_{0.5}$F$_{0.5}$BiS$_{2}$ compounds.  In the case of LaO$_{0.5}$F$_{0.5}$BiS$_{2}$, there is a strong suppression of resistivity up to $\sim$ 1.2 GPa, followed by a weaker suppression at higher pressures.  In the case of CeO$_{0.5}$F$_{0.5}$BiS$_{2}$, there is a strong suppression of resistivity up to $\sim$ 1.8 GPa, followed by a weaker suppression at higher pressures. The $\rho(P)$ data for LaO$_{0.5}$F$_{0.5}$BiS$_{2}$ and CeO$_{0.5}$F$_{0.5}$BiS$_{2}$ were taken in the normal state at 11 K and 7.6 K, respectively. These temperatures occur just above the onset of the superconducting transition at \textit{$T_c^{onset}$}. The dotted lines in Fig.~\ref{Resistivity} are guides to the eye for the rates of suppression of $\log(\rho)$ with pressure. \\

\indent There is a correlation between the pressure at which the maximum \textit{$T_c$} occurs in the \textit{$T_c(P)$} phase diagram of Fig.~\ref{Tc Combined Spline} and the pressure where the suppression rate changes in the plot of $\log(\rho)$ vs. $P$ in Fig.~\ref{Resistivity}.  For LaO$_{0.5}$F$_{0.5}$BiS$_{2}$, this ``critical pressure'' occurs at $\sim$ 1.2 GPa, while for CeO$_{0.5}$F$_{0.5}$BiS$_{2}$ it is located at $\sim$ 1.8 GPa. In both compounds, there is also an apparent pressure hysteresis as seen from the separation between the increasing pressure data (open symbols) and the decreasing pressure data (filled symbols). This pressure hysteresis becomes more pronounced where the suppression rate is higher; i.e., below the previously mentioned critical pressures.\\

The semiconducting behavior of the $\rho(T)$ data and its rapid suppression with pressure was noted in the work of Kotegawa \textit{et al.}\cite{Kotegawa} on the LaO$_{0.5}$F$_{0.5}$BiS$_{2}$ compound synthesized under high pressure. They observed that the resistivity could be described over two distinct regions by the relation
\begin{equation}\label{equation 1}
\rho(T) = \rho_0e^{\Delta/2k_{B}T}
\end{equation}
where $\rho_0$ is a constant and $\Delta$ is an energy gap. Analysis of the  $\rho(T)$ data at atmospheric pressure in these two regions, 200 - 300 K and \textit{$T_c$} - 20 K, yielded energy gaps $\Delta_1/k_B \approx$ 140 K and $\Delta_2/k_B \approx$ 1.86 K, respectively. Both energy gaps $\Delta_1$ and $\Delta_2$ were found to decrease with pressure. In this study, we have also determined values of the energy gaps $\Delta_1$ and $\Delta_2$ from linear fits of $\rho(T)$ data on a plot of log($\rho$) vs. 1/\textit{T}, as illustrated in Fig.~\ref{Lanthanum_Gap}, which displays our $\rho(T)$ data for  LaO$_{0.5}$F$_{0.5}$BiS$_{2}$ upon increasing pressure. From similar plots for LaO$_{0.5}$F$_{0.5}$BiS$_{2}$ upon decreasing pressure as well as for CeO$_{0.5}$F$_{0.5}$BiS$_{2}$ upon increasing and decreasing pressure, the two energy gaps $\Delta_1$ and $\Delta_2$, corresponding to the high and low temperature regions, respectively, could also be extracted.\\

The behavior of energy gaps $\Delta_1$ and $\Delta_2$ as a function of pressure for both LaO$_{0.5}$F$_{0.5}$BiS$_{2}$ and CeO$_{0.5}$F$_{0.5}$BiS$_{2}$ are shown in Fig.~\ref{Gap_Pressure}.  The energy gaps decrease rapidly with pressure, similar to the behavior observed by Kotegawa \textit{et al}.\cite{Kotegawa} However, it is interesting to note that the values of the energy gaps for LaO$_{0.5}$F$_{0.5}$BiS$_{2}$ at atmospheric pressure shown in Fig.~\ref{Gap_Pressure} are considerably larger than the values obtained by Kotegawa \textit{et al}.\cite{Kotegawa}  The energy gaps $\Delta_1$ and $\Delta_2$ both exhibit hysteretic behavior below a critical pressure of $\sim$ 1.3 GPa for LaO$_{0.5}$F$_{0.5}$BiS$_{2}$ and $\sim$ 2 GPa for  CeO$_{0.5}$F$_{0.5}$BiS$_{2}$; these critical pressures correlate with the pressures where the slope, d$\log(\rho)$/d$P$, changes (at temperatures in the normal state right above \textit{$T_c$}) in the $\log(\rho)$ vs. $P$ plots (Fig.~\ref{Resistivity}) and also correlate with the transition pressures into the high \textit{$T_c$} phase for both LaO$_{0.5}$F$_{0.5}$BiS$_{2}$ and CeO$_{0.5}$F$_{0.5}$BiS$_{2}$ (Fig.~\ref{Tc Combined Spline}).\\

\indent Specific heat C(\textit{T}) measurements at ambient pressure have recently been made on both the LaO$_{0.5}$F$_{0.5}$BiS$_{2}$ and CeO$_{0.5}$F$_{0.5}$BiS$_{2}$ compounds.\cite{Yazici} The samples in Ref.~\onlinecite{Yazici} and in the present study were from the same batch. In the case of LaO$_{0.5}$F$_{0.5}$BiS$_{2}$, these C(\textit{T}) measurements suggest that the superconductivity observed at $\sim$ 3 K for pressures less than $\sim$ 0.5 GPa is a bulk phenomenon. In Fig. 5 of Ref.~\onlinecite{Yazici}, there is a clear jump in C(\textit{T})/\textit{T} at \textit{$T_c$} = 2.93 K. This value of \textit{$T_c$} is close to the temperature where $\rho$ vanishes in LaO$_{0.5}$F$_{0.5}$BiS$_{2}$.  We expect, therefore, that the LaO$_{0.5}$F$_{0.5}$BiS$_{2}$ sample in this study exhibits bulk superconductivity at lower pressures.\\

It is still unclear whether or not the higher \textit{$T_c$} superconducting transitions at  pressures above 0.5 GPa are associated with bulk superconductivity. The narrow widths of the superconducting transitions would seem to suggest that the superconductivity in this pressure range is a bulk phenomenon. However, the sharpness of the resistive transitions is also consistent with a filamentary zero resistance path through the sample with a narrow distribution of \textit{$T_c$} values that could be due to small amounts of a superconducting phase, rather than bulk superconductivity.  High pressure magnetization measurements in a SQUID magnetometer were performed on several pieces of LaO$_{0.5}$F$_{0.5}$BiS$_{2}$ to determine the character of the 10 K superconducting phase by Taufour, Bud'ko and Canfield. \cite{Taufour} It was difficult to observe a diamagnetic signal against the large background from the pressure cell.  This suggests possible inhomogeneity in the LaO$_{0.5}$F$_{0.5}$BiS$_{2}$ sample.\\

In the case of CeO$_{0.5}$F$_{0.5}$BiS$_{2}$, however, there is no indication in the ambient pressure  C(\textit{T}) measurements of bulk superconductivity.\cite{Yazici} The lack of a discernible jump in specific heat, however, could be due to sample inhomogeneity and/or the proximity of \textit{$T_c$} = 1.9 K to the base temperature \textit{T} = 1.8 K of the specific heat measurements and needs to be investigated further. \\

The rapid increase of $T_c$ and broadening of the superconducting transition with pressure, as well as its reversibility with pressure for both compounds, suggest the existence of a gradual, pressure-induced transition between superconducting phases with a lower $T_c$ at lower pressure and a higher $T_c$ at higher pressure.  The broadening of the superconducting transition feature in the transition region, $\sim$1 GPa wide for both compounds, could be a consequence of the large slope of $T_c(P)$ in that pressure range (\textit{i.e.}, $\Delta T_c \simeq \left(dT_c(P)/dP\right)\Delta P$).  It might also be due to a spatial distribution of the two phases in the transition region.  In this latter scenario, as the applied pressure is increased in the transition region, the amount of the high pressure phase grows at the expense of the low pressure phase, until the sample is completely transformed into the high pressure phase at the end of the transition region. The markedly similar features in the \textit{$T_c$} versus pressure diagrams shown in Fig.~\ref{Tc Combined Spline} for the two compounds $Ln$O$_{1-x}$F$_x$BiS$_2$ (\emph{Ln} = La, Ce) suggests that this behavior could be characteristic in general of the entire class of $Ln$O$_{1-x}$F$_x$BiS$_2$ materials.\\

\indent One possible explanation for this behavior is that there is a distribution of pressures at which the transformation between the two phases takes place within the transition region.  This distribution could be associated with some type of inhomogeneity (either electronic or atomic) in the samples.  Experiments are currently in progress to search for a possible pressure-induced structural transition in these materials and to see whether the pressure-induced transition can be sharpened by improving the synthesis methods.  The synthesis of LaO$_{0.5}$F$_{0.5}$BiS$_{2}$ under pressure with a \textit{$T_c$} of $\sim$ 10 K suggests the possibility that the transformation pressure between the low and high pressure phases can be reduced to zero pressure by using a different synthesis route.\cite{Mizuguchi2}

\section{Summary}

We have observed a striking enhancement of superconductivity accompanying the suppression of semiconducting behavior with pressure in the \textit{Ln}O$_{0.5}$F$_{0.5}$BiS$_2$ compounds (\textit{Ln} = La, Ce) at critical pressures of $\sim$ 1.1 GPa and $\sim$ 2.0 GPa for \textit{Ln} = La and Ce, respectively. There is markedly similar behavior in the electrical resistivity measurements under applied pressure for these  two BiS$_2$-based superconductors LaO$_{0.5}$F$_{0.5}$BiS$_{2}$ and CeO$_{0.5}$F$_{0.5}$BiS$_{2}$. Electrical resistivity measurements reveal that for both compounds, the suppression of their semiconducting behavior is hysteretic upon application of pressure.  The semiconducting behavior of the electrical resistivity is consistent with two energy gaps that are suppressed with pressure in a similar way.  The pressure dependence of the electrical resistivity exhibits hysteresis below a critical pressure where there is a change in slope of $\log(\rho)$ vs. $P$ and where the maximum value of \textit{$T_c$} is observed.  Furthermore, for both compounds, we have discovered a continuous and reversible transient region between regions of low and high \textit{$T_c$}, which is characterized by a broadening of the superconducting transition; however, the mechanism behind the broadening of the superconducting transitions between the lower and higher \textit{$T_c$} regions is unclear. The broadening could be a simple consequence of the sensitive pressure-dependence of $T_c$ in this region, which, when coupled with even a modest pressure gradient, could result in broader measured superconducting transitions.  Sample inhomogeneity might also be responsible for the distribution of transition pressures seen in the broadening region, and the possibility of pressure-induced structural phase transitions is currently being investigated with x-ray diffraction measurements under pressure. Given the striking similarity in behavior for these two BiS$_2$-based superconductors, further electrical resistivity measurements under pressure on other compounds could point to characteristic behavior of BiS$_2$-based superconductors in general.\\

\indent In experiments currently underway we have observed the same qualitative behavior for the NdO$_{0.5}$F$_{0.5}$BiS$_2$ and PrO$_{0.5}$F$_{0.5}$BiS$_2$ compounds as were observed in the the LaO$_{0.5}$F$_{0.5}$BiS$_2$ and CeO$_{0.5}$F$_{0.5}$BiS$_2$ compounds suggesting this is indeed a general phenomenon in the class of \textit{Ln}O$_{0.5}$F$_{0.5}$BiS$_2$ compounds. Our results on the NdO$_{0.5}$F$_{0.5}$BiS$_2$ compound may be compared to the recently reported study of NdO$_{0.5}$F$_{0.5}$BiS$_{2}$ specimens prepared in a solid state reaction by Selvan \textit{et al}.\cite{Selvan} For the  compounds with \textit{Ln} = La, Ce and Pr, there is a dramatic decrease in the electrical resistivity with pressure that reflects a continuous suppression of semiconducting behavior. Although the temperature coefficient of the electrical resistivity, $d\rho/dT$, at the highest pressures is small for these \textit{Ln} = La, Ce and Pr compounds, the coefficient nevertheless remains negative ($d\rho/dT < $ 0), so that we cannot definitely conclude that the metallic state has been achieved. We have, however, been able to reach a metallic state for \textit{Ln} = Nd, indicated by a positive temperature coefficient of resistivity ($d\rho/dT > $ 0), consistent with a semiconductor-metal transition. Experiments to pressures in excess of 3 GPa are currently under way to see if definitive metallic states (i.e.,  $d\rho/dT > $ 0) can be attained for the \textit{Ln} = La, Ce, and Pr compounds.

\begin{acknowledgements}

High pressure research at the University of California, San Diego (UCSD) was supported by the National Nuclear Security Administration under the Stewardship Science Academic Alliance Program through the U.S. Department of Energy (DOE) under Grant No. DE-52-09NA29459.  Sample synthesis at UCSD was sponsored by the U.S. Air Force Office of Scientific Research under MURI Grant No. FA9550-09-1-0603 (``Broad-based search for new and practical superconductors,'' involving the University of Maryland, Iowa State University, and the University of California, San Diego).   Characterization of samples at ambient pressure was supported by the U.S. DOE Grant No. DE-FG02-04-ER46105. We would like to thank V. Taufour, S. L. Bud'ko, and P. C. Canfield of Iowa State University for performing magnetization measurements under pressure on several LaO$_{1-x}$F$_x$BiS$_2$ samples  under the auspices of the above mentioned AFOSR MURI project.

\end{acknowledgements}

\bibliography{References_La_Ce}

\end{document}